\documentclass[twoside,a4paper,11pt]{sea10}
\usepackage{graphicx}
\usepackage{hyperref}
\usepackage{movie15}

\newcommand{\wmap}{{\it WMAP }}
\newcommand{\planck}{{\it Planck }}

\def\reff@jnl#1{{\rm#1\/}}
\def\aj{\reff@jnl{AJ}}                  
\def\araa{\reff@jnl{ARA\&A}}            
\def\apj{\reff@jnl{ApJ}}                
\def\apjl{\reff@jnl{ApJ}}               
\def\apjs{\reff@jnl{ApJS}}              
\def\ao{\reff@jnl{Appl.Optics}}         
\def\apss{\reff@jnl{Ap\&SS}}            
\def\aap{\reff@jnl{A\&A}}               
\def\aapr{\reff@jnl{A\&A~Rev.}}         
\def\aaps{\reff@jnl{A\&AS}}             
\def\azh{\reff@jnl{AZh}}                
\def\baas{\reff@jnl{BAAS}}              
\def\gca{\reff@jnl{GeCoA}}              
\def\jrasc{\reff@jnl{JRASC}}            
\def\memras{\reff@jnl{MmRAS}}           
\def\mnras{\reff@jnl{MNRAS}}            
\def\pra{\reff@jnl{Phys.Rev.A}}         
\def\prb{\reff@jnl{Phys.Rev.B}}         
\def\prc{\reff@jnl{Phys.Rev.C}}         
\def\prd{\reff@jnl{Phys.Rev.D}}         
\def\prl{\reff@jnl{Phys.Rev.Lett}}      
\def\pasp{\reff@jnl{PASP}}              
\def\pasj{\reff@jnl{PASJ}}              
\def\qjras{\reff@jnl{QJRAS}}            
\def\skytel{\reff@jnl{S\&T}}            
\def\solphys{\reff@jnl{Solar~Phys.}}    
\def\sovast{\reff@jnl{Soviet~Ast.}}     
\def\ssr{\reff@jnl{Space~Sci.Rev.}}     
\def\zap{\reff@jnl{ZAp}}                
\def\nat{\reff@jnl{Nature}}             
\def\procspie{\reff@jnl{SPIE Conference Series}}             

\begin{document}
\pagenumbering{arabic}
\pagestyle{myheadings}
\thispagestyle{empty}
{\flushleft\includegraphics[width=\textwidth,bb=58 650 590 680]{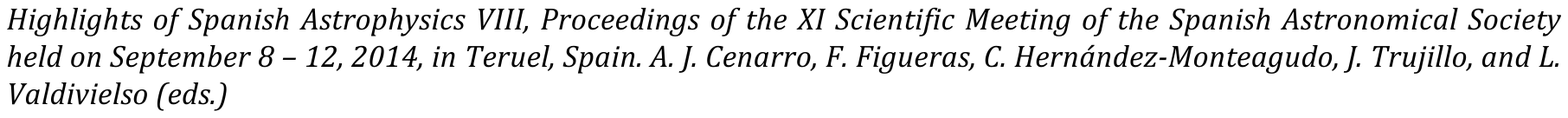}}
\vspace*{0.2cm}
\begin{flushleft}
{\bf {\LARGE
The QUIJOTE experiment: project overview and first results
}\\
\vspace*{1cm}
R. G\'enova-Santos$^{1,6}$, J.A. Rubi\~no-Mart\'{\i}n$^{1,6}$, R. Rebolo$^{1,6,7}$, M. Aguiar$^1$, F. G\'omez-Re\~nasco$^1$, C. Guti\'errez$^{1,6}$, R.J. Hoyland$^1$, C. L\'opez-Caraballo$^{1,6,8}$, A.E. Pel\'aez-Santos$^{1,6}$, 
M.R. P\'erez-de-Taoro$^1$, F. Poidevin$^{1,6}$ , V. S\'anchez de la Rosa$^{1}$, D. Tramonte$^{1,6}$, A. Vega-Moreno$^{1}$, T. Viera-Curbelo$^{1}$, R. Vignaga$^{1,6}$, 
E. Mart\'{\i}nez-Gonz\'alez$^2$, R.B. Barreiro$^2$, B. Casaponsa$^2$, F.J. Casas$^2$, J.M. Diego$^2$, R. Fern\'andez-Cobos$^2$, D. Herranz$^2$, M. L\'opez-Caniego$^2$, D. Ortiz$^2$, 
P. Vielva$^2$, E. Artal$^3$, B. Aja$^3$, J. Cagigas$^3$, J.L. Cano$^3$, L. de la Fuente$^3$, A. Mediavilla$^3$, J.V. Ter\'an$^3$, E. Villa$^3$, L. Piccirillo$^4$, 
Davies$^4$, R.J. Davis$^4$, C. Dickinson$^4$, K. Grainge$^4$, S. Harper$^4$, B. Maffei$^4$, M. McCulloch$^4$, S. Melhuish$^4$, G. Pisano$^4$,
R.A. Watson$^4$, A. Lasenby$^{5,9}$, M. Ashdown$^{5,9}$, M. Hobson$^5$, Y. Perrott$^5$, N. Razavi-Ghods$^5$, R. Saunders$^6$, D. Titterington$^6$ and P. Scott$^6$
}\\
\vspace*{0.5cm}
$^1$ Instituto de Astrofis\'{i}ca de Canarias, 38200 La Laguna, Tenerife, Canary Islands, Spain\\
$^2$ Instituto de F\'{\i}sica de Cantabria (CSIC-Universidad de Cantabria), Avda. de los Castros s/n, 39005 Santander, Spain\\
$^3$ Departamento de Ingenieria de COMunicaciones (DICOM), Laboratorios de I+D de Telecomunicaciones, Universidad de Cantabria, Plaza de la Ciencia s/n, E-39005 Santander, Spain\\
$^4$ Jodrell Bank Centre for Astrophysics, Alan Turing Building, School of Physics and Astronomy, The University of Manchester, Oxford Road, Manchester, M13 9PL, U.K\\
$^5$ Astrophysics Group, Cavendish Laboratory, University of Cambridge, J.J. Thomson Avenue, Cambridge CB3 0HE, UK\\
$^6$ Departamento de Astrof\'{\i}sica, Universidad de La Laguna (ULL), 38206 La Laguna, Tenerife, Spain\\
$^7$ Consejo Superior de Investigaciones Cient\'{\i}ficas, Spain\\
$^8$ Departamento de F\'{\i}sica, Universidad de la Serena, Av. Cisternas 1200, La Serena, Chile\\
$^9$ Kavli Institute for Cosmology, Madingley Road, Cambridge, CB3 0HA
\end{flushleft}

\markboth{
The QUIJOTE experiment 
}{ 
G\'enova-Santos et al.
}
\thispagestyle{empty}

\vspace*{0.4cm}
\begin{minipage}[l]{0.09\textwidth}
\ 
\end{minipage}
\begin{minipage}[r]{0.9\textwidth}
\vspace{1cm}
\section*{Abstract}{\small
QUIJOTE (Q-U-I JOint TEnerife) is a new polarimeter aimed to characterize the polarization of the Cosmic Microwave Background 
and other Galactic and extragalactic signals at medium and large angular scales in the frequency range 10-40 GHz.
The multi-frequency (10-20~GHz) instrument, mounted on the first QUIJOTE telescope, saw first light on November 2012 
from the Teide Observatory (2400~m a.s.l). During 2014 the second telescope has been installed at this observatory. A second 
instrument at 30~GHz will be ready for commissioning at this telescope during summer 2015, and a third additional instrument 
at 40~GHz is now being developed. These instruments will have nominal sensitivities to detect the B-mode polarization due to 
the primordial gravitational-wave component if the tensor-to-scalar ratio is larger than $r=0.05$.
\normalsize}
\end{minipage}

\section{Introduction \label{sec:intro}}
The Cosmic Microwave Background (CMB) is recognised as one of the most powerful cosmological probes. The study of temperature 
anisotropies by missions like \wmap \cite{bennett13} or \planck \cite{cpp1}, and previous ground-based and balloon-born experiments, 
have reached levels of sensitivity and angular resolution that have allowed the determination of the main cosmological parameters with 
accuracies close to 1\%. The CMB polarization anisotropies also encodes a wealth of cosmological information. Not only the E-mode 
polarization have allowed to tighten the cosmological constraints by breaking degeneracies between parameters, but the detection 
of the B-mode polarization may provide a confirmation of the existence of primordial gravitational waves created by inflation, the 
epoch of exponential expansion in the primordial Universe \cite{kamionkowsky97, zaldarriaga97}. Specifically-targeted experiments like
QUIET \cite{quiet12} or BICEP \cite{bicep114} have already started to put constraints on the tensor-to-scalar ratio, $r$, the parameter that 
is used to parameterise the amplitude of the B-mode signal. Others like SPTpol \cite{hanson13} or POLARBEAR \cite{polarbear14} have 
measured the small-angular scale B-mode component which is not primordial but originated by the lensing of the E-mode polarization by 
large-scale-structure. Recently, the first primordial B-mode detection was claimed in data from the BICEP2 experiment~at 150~GHz, with 
a level of the tensor-to-scalar ratio $r=0.20^{+0.07}_{-0.05}$ \cite{bicep214}. However, data from the \planck satellite have shown that 
the level of polarized dust emission in the region of the sky covered by BICEP2 could form a significant component of the measured 
signal \cite{pip30}, thus causing a likely reduction in the level of cosmological signal that can be inferred from the BICEP2 results.

It is well recognised that any unambiguous detection of the B-mode anisotropy requires a detailed assessment of the level of foreground 
contamination and ideally confirmation by independent experiments operating at different frequencies. The QUIJOTE 
(Q-U-I JOint TEnerife) experiment \cite{quijote12,quijote14}, thanks to its wide frequency coverage (10-40~GHz), will provide the 
characterization of the polarization of the synchrotron and anomalous microwave emission (AME), and of the B-mode signal down to a 
sensitivity of $r=0.05$. Updated information of the project can be found in {\tt http://www.iac.es/project/cmb/quijote}.

\section{Project baseline\label{project}}

The QUIJOTE experiment is a scientific collaboration between the Instituto de Astrof\'{\i}sica de Canarias, the Instituto de F\'{\i}sica de 
Cantabria, the IDOM company, and the universities of Cantabria, Manchester and Cambridge. The project consists of two 
telescopes and three instruments covering the frequency range $10-40$~GHz, with an angular resolution of 
$\sim 1^\circ$, and located at the Teide observatory (2400~m) in Tenerife (Spain). This site provides excellent atmospheric conditions 
for CMB observations, as demonstrated by previous experiments (Tenerife experiment, IAC-Bartol, JBO-IAC interferometer, 
COSMOSOMAS, VSA). Data obtained with the first QUIJOTE experiment throughout one year shows that the zenith atmosphere 
temperature is on average $\sim 2~$K at 11~GHz and $\sim 4 - 6~$K at 19~GHz, while the PWV column density is typically between 2 
and 4~mm. 

\begin{figure}
\center
\includegraphics[height=4.9cm]{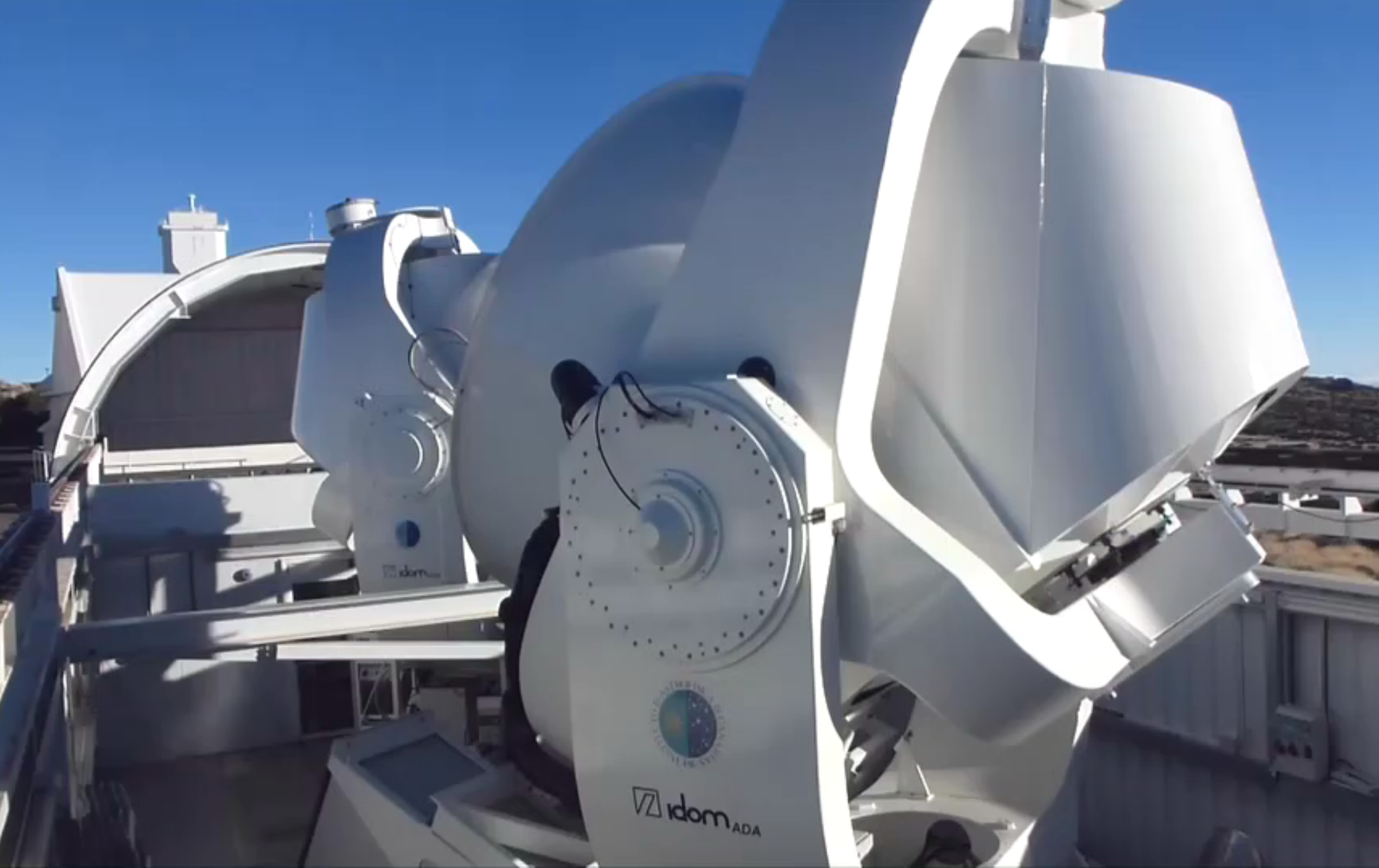} 
\includegraphics[height=4.9cm]{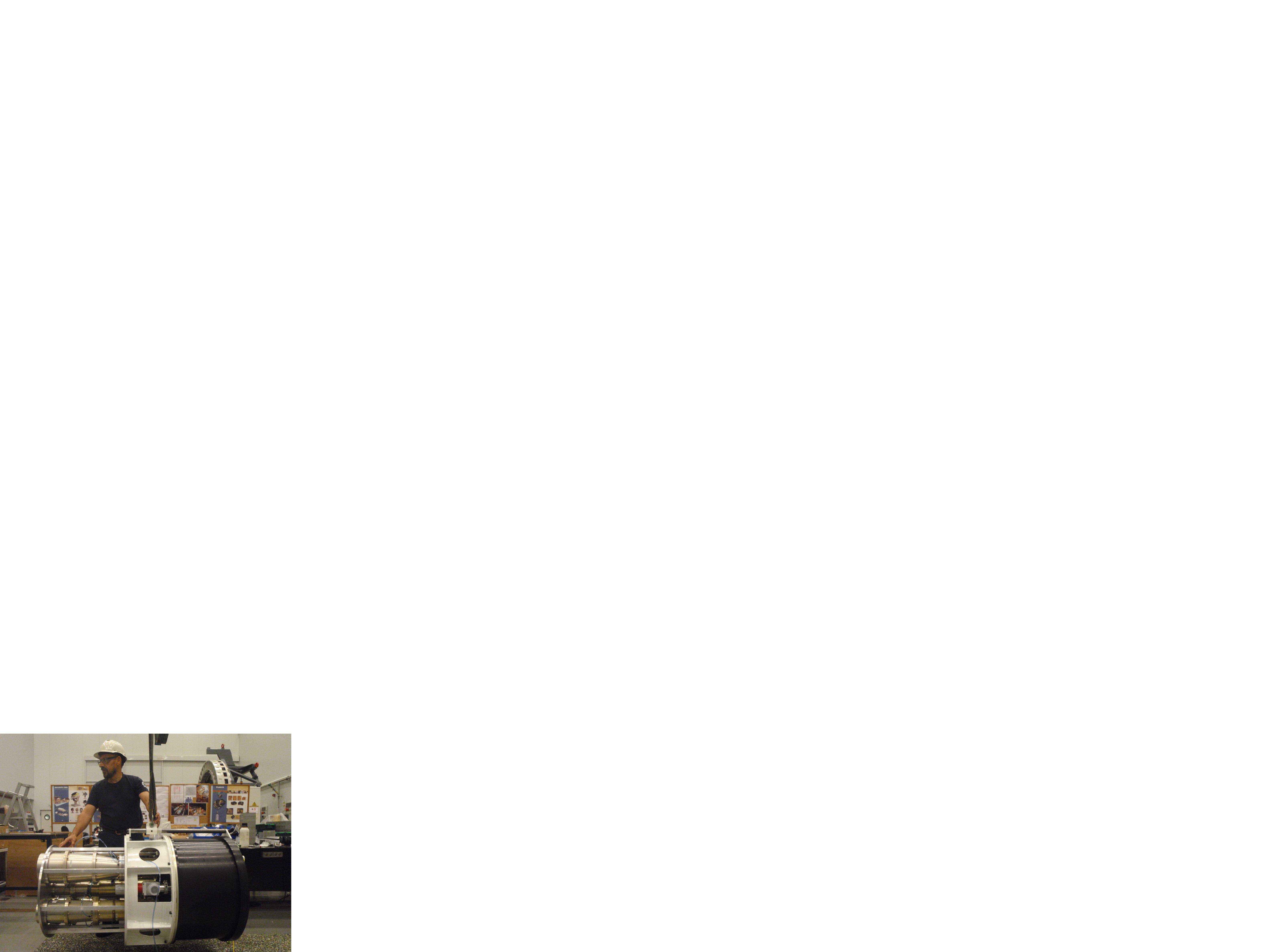}
\caption{\label{fig:qt1_qt2_mfi} Left: QT1 (background) and QT2 (front) inside the enclosure, at the Teide observatory. Right: MFI, during 
integration tests (December, 2011).
}
\end{figure}

The first QUIJOTE telescope (QT1) is currently fitted with the multi-frequency instrument (MFI), which has four frequency bands centred 
in 11, 13, 17 and 19~GHz, respectively. It saw first light on November 2012, and ever since is performing routine observations of different 
fields. Some results obtained with this experiment will be presented in Section~\ref{sec:science}. The second QUIJOTE telescope (QT2) 
was installed at the observatory on July 2014. This telescope will be fitted with the thirty-gigahertz instrument (TGI), consisting in 
31 polarimeters at 30~GHz and which will start commissioning on April 2015. A third set of detectors, the forty-gigahertz instrument 
(FGI) is being constructed at the time of writing (January 2015). Figure~\ref{fig:qt1_qt2_mfi} shows photos of QT1, QT2 and the MFI.

In Table~\ref{tab:characteristics} we show the nominal characteristics of the three QUIJOTE experiments. The noise equivalent power 
for each frequency band is computed as:
\begin{equation}
{\rm NEP}_{\rm MFI} = \frac{T_{\rm sys}}{\sqrt{\Delta\nu}}~~,~~
{\rm NEP}_{\rm TGI,FGI} = \sqrt{2}\frac{T_{\rm sys}}{\sqrt{\Delta\nu N_{\rm chan}}}~~,
\label{eq:nep}
\end{equation}
where $T_{\rm sys}$ stands for the total system temperature, $\Delta\nu$ is the bandwidth and $N_{\rm chan}$ the number of channels 
(computed here as the number of horns times the number of output channels per horn). The NEPs are different for MFI compared to TGI 
and FGI because of their different strategies for measuring the polarization. 
In the MFI it involves differentiating pairs of channels, whereas the TGI and the FGI will make use of electronic modulation providing an 
instantaneous measurement of $Q$ and $U$ for each channel. The MFI parameters measured using real observations are in good 
agreement with the nominal parameters shown in Table~\ref{tab:characteristics}. In particular, the measured $Q$ and $U$ NEPs are in 
the range $644-792~\mu$K~$s^{1/2}$ for different frequency channels. In total intensity $I$, where the instrument knee frequencies 
are worse ($f_{\rm k}\sim 0.1-1$~Hz for $Q$ and $U$ and $f_{\rm k}\sim 10-100$~Hz for $I$) the instantaneous sensitivities 
are typically a factor $\sim 2.5$ worse. 

\begin{table}[ht]
\caption{Nominal characteristics of the three QUIJOTE instruments: MFI, TGI and FGI. Sensitivities are referred to Stokes $Q$ and $U$ 
parameters.}
\center
\begin{minipage}{0.5\textwidth}
\center
\begin{tabular}{lcccccc}
\hline\hline
& \multicolumn{4}{c}{MFI} & TGI & FGI \\
\hline
Nominal frequency [GHz] & 11 & 13 & 17 & 19 & 30 & 40 \\
Bandwidth [GHz] & 2 & 2 & 2 & 2 & 10 & 12\\
Number of horns & 2 & 2 & 2 & 2 & 31 & 31\\
Channels per horn & 4 & 4 & 4& 4 & 4 & 4 \\
Beam FWHM ($^\circ$) & $0.92$ & $0.92$ & $0.60$ & $0.60$ & $0.37$ & $0.28$ \\
$T_{\rm sys}$ [K] & 25 & 25 & 25 & 25 & 35 & 45 \\
NEP [$\mu$K~$s^{1/2}$] & 559 & 559 & 559 & 559 & 44 & 52 \\
Sensitivity [Jy~$s^{1/2}$] &  $0.61$ & $0.85$ & $0.62$ & $0.77$ & $0.06$ & $0.07$ \\ 
\hline
\end{tabular}
\end{minipage}
\label{tab:characteristics}
\end{table}

\section{Experimental details\label{sec:experiment}}

In this section we will present a succinct description of the technical details of the project. More extended details can be found 
in previous proceedings \cite{gomez10,mfi,tgi,gomez12}, and in more extended papers that are in preparation.

\subsection{Telescopes}
Both the QT1 and the QT2 are based on an offset crossed-Dragone optic, with projected apertures of $2.25$~m and $1.89$~m for the 
primary (parabolic) and secondary (hyperbolic) mirrors, and provide highly symmetric beams (measured ellipticity $> 0.98$) with very 
low sidelobes ($\leq -40$~dB) and polarization leakage ($\leq -25$~dB). These mirrors are supported by an altazimuth mount, which 
allows rotation around the vertical azimuth axis at a maximum speed of 6~rpm ($36^\circ$/s). The QT1 and QT2 mirrors have been 
manufactured with surface accuracies to make them operative up to 90~GHz and 200~GHz, respectively.

\subsection{Multi-frequency Instrument (MFI)}
This instrument is fed with four independent sky pixels: two of them operate at $10-14$~GHz and the other two at $16-20$~GHz. 
Each pixel consists in a conical corrugated feedhorn feeding a novel cryogenic on-axis stepping polar modulator. Continuous rotation 
of these polar modulators, which would allow instantaneous measurements of $I$, $Q$ and $U$, is not appropriate for long-term 
operation of the system, and instead we obtain $Q$ and $U$ by differentiating pairs of channels, and step each modulator with a 
periodicity of $\sim$~one day in order to reduce systematics. The input orthogonal linear polar signals are separated by a wide-band 
cryogenic OMT (ortho-mode-transducer) and later amplified through a pair of MMIC (monolithic microwave integrated circuit) 
6-20~GHz LNAs (low-noise-amplifiers). These signals are then fed into a room-temperature Back-End module (BEM), where they 
are further amplified, and later split and passed through a $180^\circ$-hybrid, providing four outputs. Each of these outputs is then 
split and the total band passes spectrally filtered into an upper and lower band, each with a bandwidth of 2~GHz. Therefore, each 
pixel provides a total of eight channels, four in each of these two bands (see Table~\ref{tab:characteristics}). The low-frequency pixels 
provide channels centred in 11 and 13~GHz, and the high-frequency pixels centred in 17 and 19~GHz. In practice, only half of 
these channels are used to measure $Q$ and $U$, as the pair differences result in the removal of the $1/f$ noise. The other are 
affected by different $1/f$ noises as the two channels of each pair pass through different LNAs, and are not used. We plan however 
to correct for this by the installation of two $90^\circ$-hybrids, after which the eight channels will be useful and as a consequence 
the MFI sensitivity of equation~\ref{eq:nep} will be improved by a factor 2.

\subsection{Thirty-gigahertz Instrument (TGI)}
The TGI will be fitted with 31 polarimeters sensitive in the frequency range $26-36$~GHz. As it was said before, the original design of 
the MFI, based on spinning polar modulators, was found to be not suitable for long-term operations. Thus, in the TGI we have modified 
the receiver configuration by replacing the polar modulators by a fixed polarizer. The modulation of the polarized signal is achieved by 
the combination of one $90^\circ$ and one $180^\circ$ phase switch. The two states of these switches are exchanged in order to 
generate four polarization states to minimise the different systematics of the receiver.

\subsection{Forty-gigahertz Instrument (FGI)}
The FGI will be fitted with 31 polarimeters working in the frequency range $35-47$~GHz. The conceptual design is the same as in the TGI.

\section{Science cases and first results\label{sec:science}}

\subsection{Core science}

The QUIJOTE project is envisaged to achieve two primary scientific goals:
\begin{itemize}{\parsep=-1mm}
\item to detect the B-mode signal from primordial gravitational waves down to a sensitivity $r=0.5$;
\item to determine the polarization properties of the synchrotron and anomalous microwave emissions from our Galaxy 
at low frequencies ($10-40$~GHz).
\end{itemize}
To meet these goals we will perform two polarization surveys:
\begin{itemize}{\parsep=-1mm}
\item[i)] a wide Galactic survey. It covers around $20,000$~deg$^2$ and after 6 months of observations with each instrument 
it will have a final sensitivity of $\sim 25~\mu$K/beam at 11, 13, 17 and 19~GHz (MFI), $\sim 4~\mu$K/beam at 30~GHz (TGI) and
$\sim 6~\mu$K/beam at 40~GHz (FGI). Currently we are finalising this survey with the MFI, and have accumulated 5.5 months of data;
\item[ii)] a deep ``cosmological'' survey. It will encompass around $3,000$~deg$^2$ . Here we shall obtain a sensitivity of $\sim 5~\mu$K/beam after 2 years of observations with the MFI (11-19~GHz), and $\sim 1~\mu$K/beam with the TGI (30~GHz) and with the FGI (40~GHz).
\end{itemize}

According to these nominal sensitivities, QUIJOTE will provide one of the most sensitive measurements of the polarization of the 
synchrotron and anomalous microwave emissions in the frequency rage 10-20~GHz. This is essential as the B-mode signal is known to be
sub-dominant over the Galactic synchrotron in our frequency range. Using the MFI maps from the deep survey we will be able to determine 
the amplitude and spectral index of the synchrotron emission in every individual pixel, and therefore extrapolate to 30 and 40~GHz in 
order to correct the TGI and FGI maps. Our goal is to have a residual synchrotron emission at these frequencies below the noise 
sensitivity of these maps. According to the forecasted sensitivities in the deep cosmological survey, we have shown in a previous 
publication\cite{quijote12} that after 1~year of effective observing time over $3,000$~deg$^2$ with the TGI we could reach a sensitivity 
on the tensor-to-scalar ratio of $r=0.1$ (at the 95\% C.L.). The combination of 3~years of effective time with the TGI and 2~years with the 
FGI (we can observe simultaneously with the two instruments because we have two telescopes) would allow to reach $r=0.05$.

\subsection{Non-core science}

\begin{figure}
\center
\includegraphics[height=6.7cm]{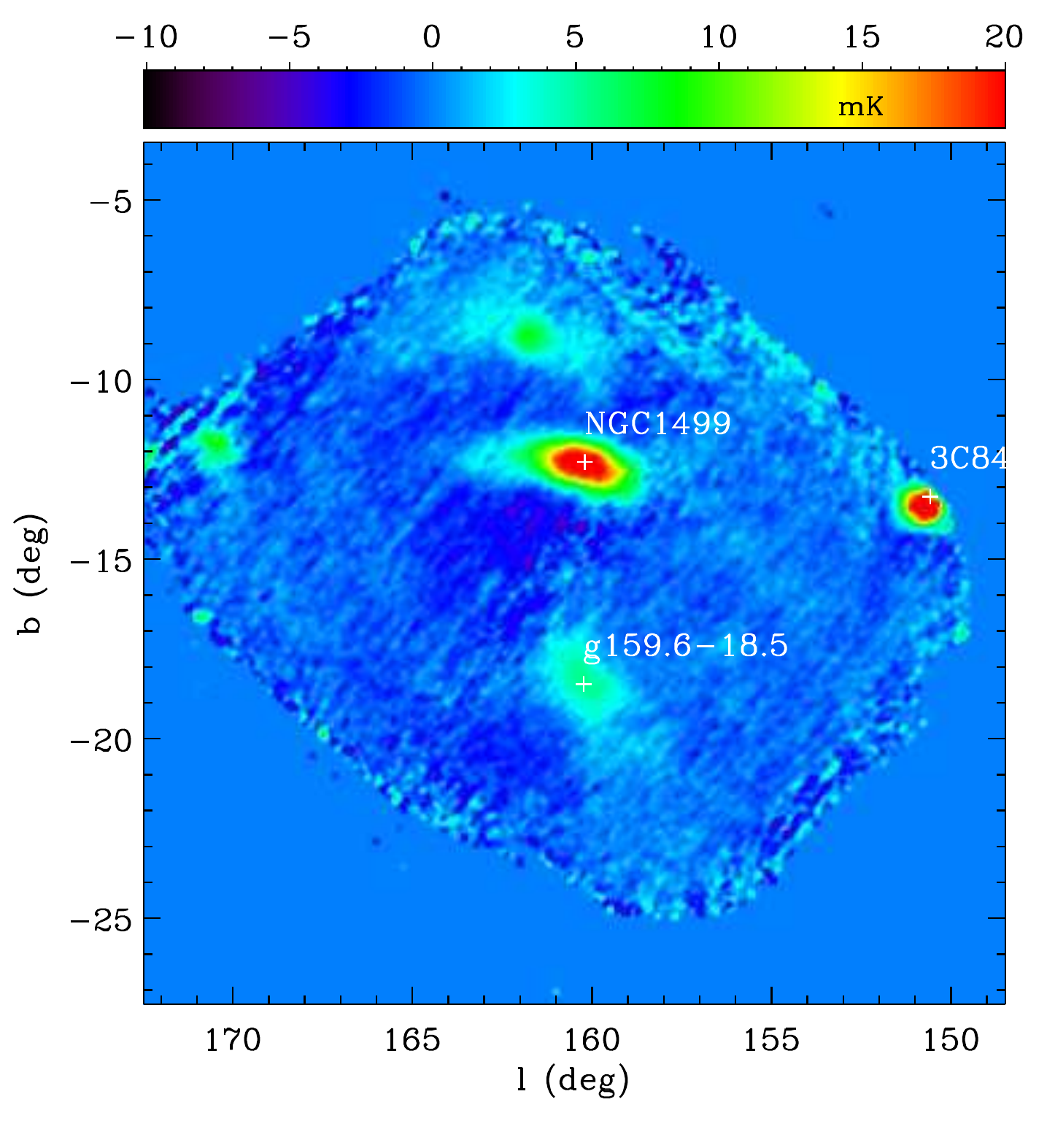} 
\includegraphics[height=6.5cm]{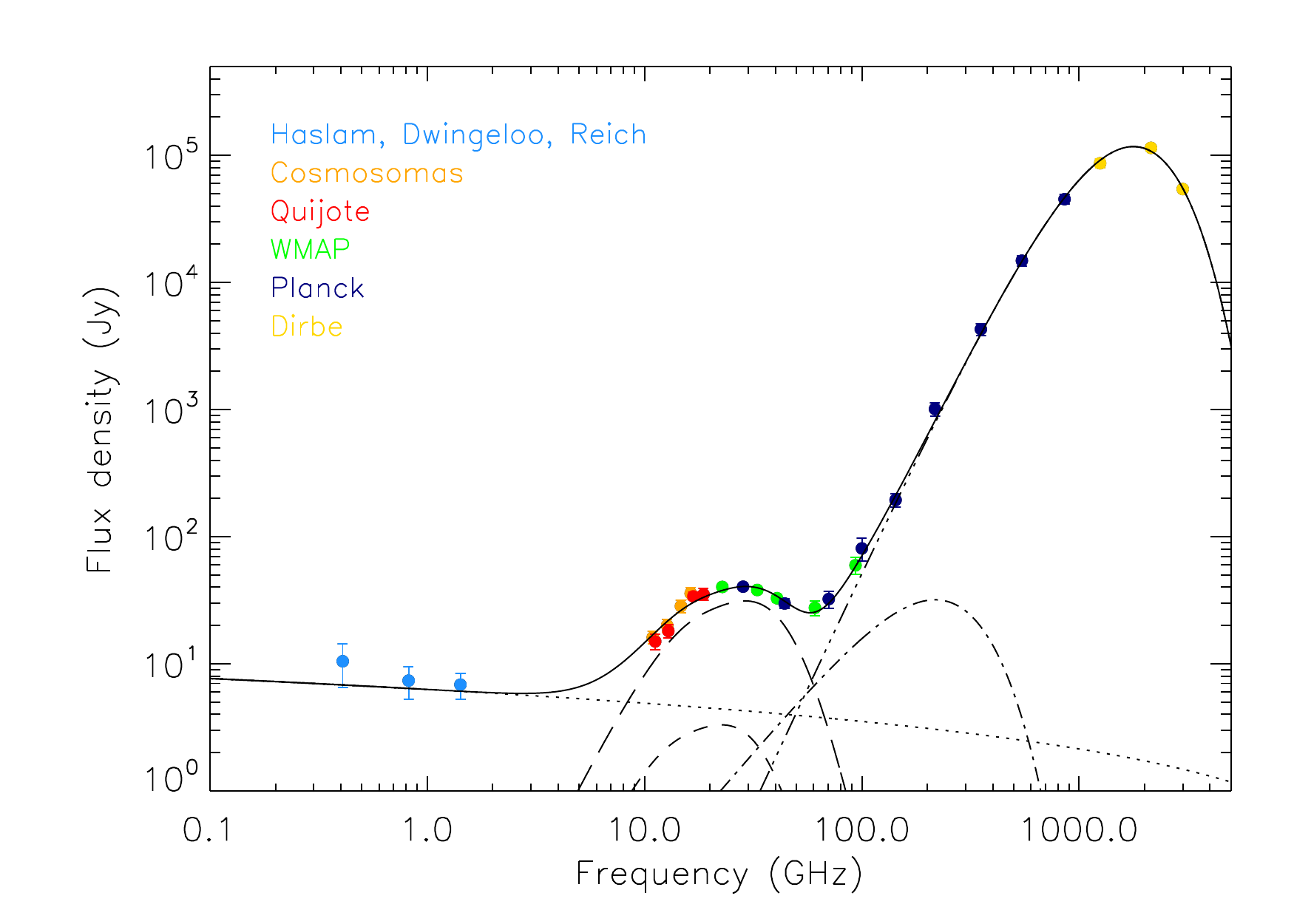}
\caption{\label{fig:map_sed_perseus} Left: QUIJOTE intensity map at 11 GHz of the region around the Perseus molecular complex 
(G159.6-18.5). The California nebula (NGC1499) and the quasar 3C84 are also visible. Right: spectral energy distribution of G159.6-18.5. 
The observed data points are fitted to a combination of free-free emission, two spinning dust components, CMB and thermal dust.
}
\end{figure}

So far, we have invested a significant amount of time to observe different fields related with non-core since programmes. Some of these are:\\

i) Study of the polarization of the Anomalous Microwave Emission (AME). Currently there is very limited observational information about the level 
of polarization of the AME, and 
only upper limits, typically at $<1\%$ \cite{lopez11,dickinson11}, have been published. We have dedicated 149 hours to observe a 250~deg$^2$-region 
around the Perseus molecular complex, and have derived upper limits on the AME from G159.6-18.5 of $<6.3\%$ and $<2.8\%$  
respectively at 12 and 18~GHz, a spectral range that had not been covered before in polarization. In Figure~\ref{fig:map_sed_perseus} 
we show the 11~GHz intensity map around this region, and the spectral energy distribution of G159.6-18.5, which represents 
the  most-precise AME spectrum measured to-date, with 13 points begin dominated by AME. These results have been included in a paper 
that we have recently submitted to MNRAS \cite{genova15}. We have new observations on G159.6-18.5, amounting to 465 hours, and over a smaller 
area of $\sim 30$~deg$^2$, which could potentially lead to upper limits better by a factor $\sim 5$.\\

\begin{figure}
\center
\includegraphics[width=16.5cm]{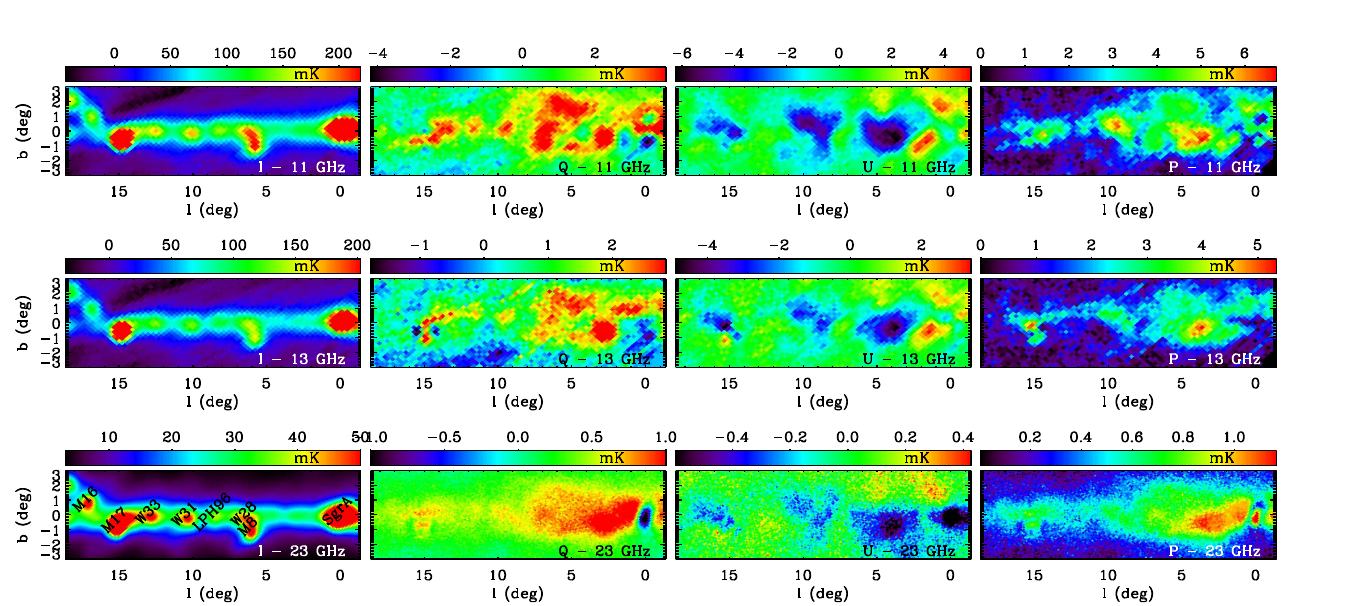} 
\caption{\label{fig:map_haze} QUIJOTE and WMAP maps along the Galactic centre. We show QUIJOTE maps at 11 and 13~GHz and 
WMAP maps at 23~GHz. From left to right maps correspond respectively to total intensity Stokes $I$, $Q$ and $U$ and polarized intensity 
$P$. In the WMAP $I$ map we indicate the names of some of the detected regions.
}
\end{figure}

ii) Study of the WMAP haze in polarization. This is a region with an excess of microwave emission in the region around the Galactic centre, 
initially found in WMAP data, with a spectrum significantly flatter than synchrotron, and which was later discovered to have a 
$\gamma$-ray counterpart in Fermi data \cite{dobler12}. There is currently a strong debate about its origin. One appealing hypothesis is  
based on hard  synchrotron radiation driven by relativistic electrons and positrons produced in the annihilations of one (or more) species of 
dark matter particles. QUIJOTE data could have an important contribution here, as it could allow us to measure or to constrain the 
expected level of polarization of this synchrotron emission. So far we have accumulated 406~hours of data in a $\sim 700$~deg$^2$ 
region around the Galactic centre. In Figure~\ref{fig:map_haze} we show QUIJOTE maps at 11 and 13~GHz resulting from 97~hours of 
data, in comparison with WMAP maps at 23~GHz. A clear correlation can be seen between the polarized structures probed by 
QUIJOTE and WMAP. We are preparing a paper where we will present the results  inferred from these observations, and their implications 
on the physical origin of the haze.\\

\begin{figure}
\center
\includegraphics[width=12cm]{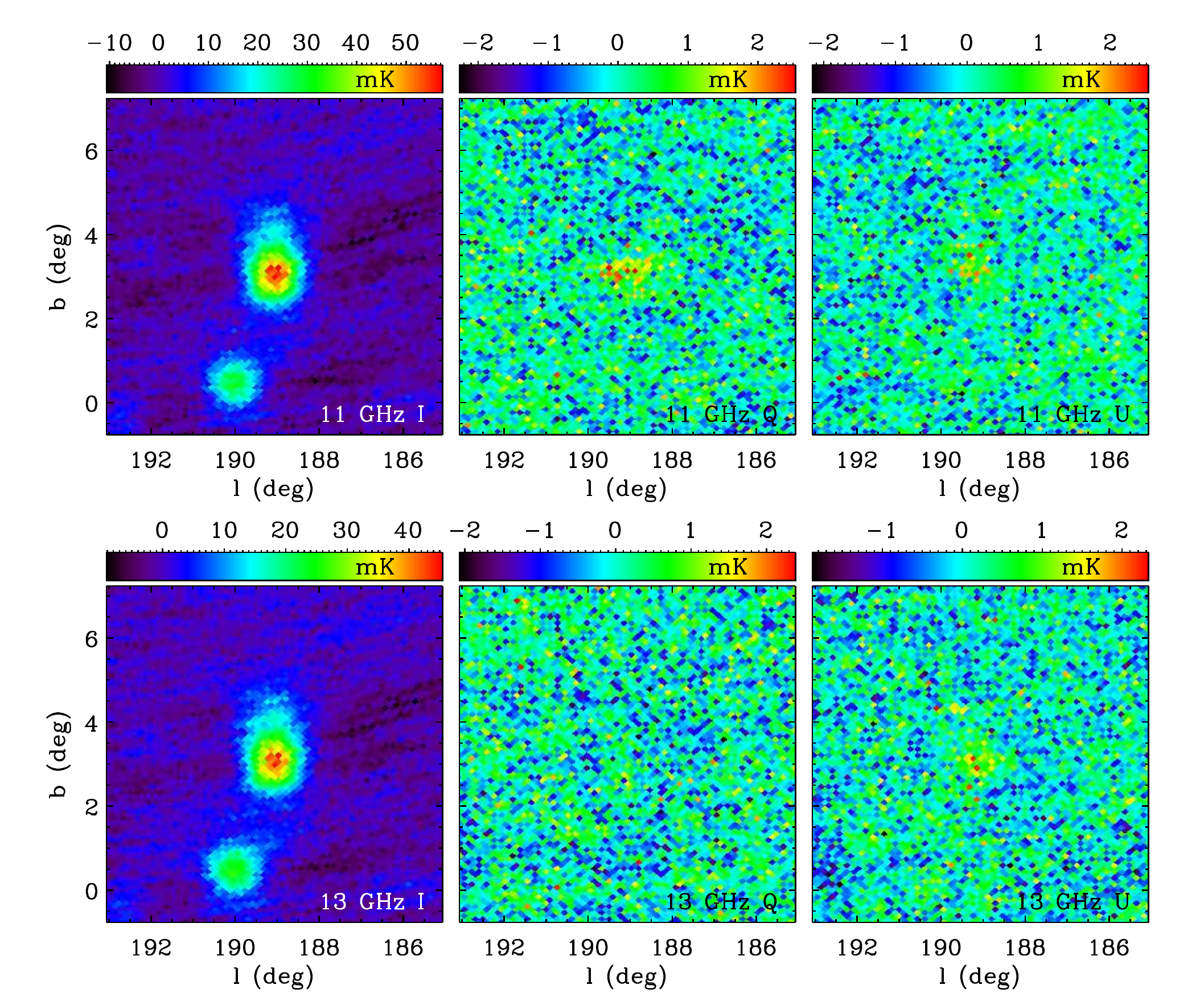} 
\caption{\label{fig:maps_ic443} QUIJOTE 11 and 13~GHz maps around the SNR IC443. Polarized emission is marginally seen, with 
polarization angle $\gamma = -49.8\pm 19.5^\circ$ and $-55.7\pm 12.0^\circ$, respectively at 11 and 13~GHz, values that are compatible 
with WMAP 23 GHz, $\gamma_{\rm WMAP}=-31.7\pm 5.7^\circ$. The amount of data used to build these maps correspond to 31~hours.
}
\end{figure}

iii) Fan region. The Fan region is one of the brightest features of the polarized radio continuum sky, located around $l=140^\circ$, 
$b=6^\circ$, and long thought to be due to local ($d<500$~pc) synchrotron emission. This region is an interesting test-bench to 
assess the potentiality of QUIJOTE to recover diffuse polarized emission. At the time of writing we have accumulated 251~hours of data 
on a $\sim 380$~deg$^2$ region, covering not only the diffuse emission but the point-like emission from the 3C58 SNR.\\

iv) Study of SNRs. We are interested in the analysis of the spectral energy distributions of SNRs, in order to analyse possible curvatures of
the synchrotron spectrum. The wide-survey, which covers the full northern sky, will have enough sensitivity to study different Galactic SNRs. 
A higher sensitivity is achieved in 3C58 in the Fan observations. We also observe, practically on a daily basis, Tau A, which is our primary 
calibrator. At the moment we have in total 204~hours in this source, distributed in 631 individual raster scans, each of $\approx 20$~min. 
These observations can be used to study the secular decrease of Tau at the QUIJOTE frequencies. We have also collected 44~hours in 
IC443 and 75~hours in W63. In Figure~\ref{fig:maps_ic443} we show 11 and 13~GHz maps on IC443, built from 31~hours of data.\\

v) Study of the polarization properties of point sources. Apart from the previous SNRs, most of which are point sources, we have observed other Galactic and extra-galactic point sources. In particular: Cas A, which we also use as a calibrator, 3C273, NGC7027 and 3C286. Also, we have been recently awarded 4~hours of VLA time in the semester 2015A, to observe at 30 and 40~GHz pre-selected sources in the 
deep-survey fields.\\

\small 

\section*{Acknowledgments}   
The QUIJOTE-CMB experiment is being developed by the Instituto de Astrof\'{\i}sica de Canarias (IAC), the Instituto de F\'{\i}sica de Cantabria (IFCA), and the Universities of Cantabria, Manchester and Cambridge. Partial financial support is provided by the Spanish Ministry of Economy and Competitiveness (MINECO) under the projects AYA2007-68058-C03-01, AYA2010-21766-C03-02, and also by the Consolider-Ingenio project CSD2010-00064 (EPI: Exploring the Physics of Inflation).


\begin{thebibliography}{}
\small
\bibitem{bennett13} Bennett, C.~L., Larson,
D., Weiland, J.~L., et al.\ 2013, \apjs, 208, 20
\bibitem{bicep114} BICEP1 Collaboration,
Barkats, D., Aikin, R., Bischoff, C., et al.\ 2014, \apj, 783, 67
\bibitem{bicep214} BICEP2 Collaboration, Ade, P.~A.~R., Aikin,
R.~W., Barkats, D., et al.\ 2014, Physical Review Letters, 112, 241101
\bibitem{dickinson11} Dickinson, C., Peel, M., \& Vidal, M.\ 2011, \mnras, 418, L35 
\bibitem{dobler12} Dobler, G.\ 2012, \apj, 750, 17 
\bibitem{genova15}  G{\'e}nova-Santos, R., Rubi{\~n}o-Mart{\'{\i}}n, J.~A., Rebolo, R., et al.\ 2015, arXiv:1501.04491 
\bibitem{gomez10} Gomez, A., Murga, G., Etxeita, B., et al.\ 2010, \procspie, 7733, 77330Z 
\bibitem{gomez12}  G{\'o}mez-Re{\~n}asco, M.~F., Aguiar, M., Herreros, J.~M., et al.\ 2012, \procspie, 8452, 845234 
\bibitem{hanson13} Hanson, D., Hoover, S., 
Crites, A., et al.\ 2013, Physical Review Letters, 111, 141301 
\bibitem{mfi} Hoyland, R.~J., Aguiar-Gonz{\'a}lez, M., Aja, B., et al.\ 2012, \procspie, 8452, 845233 
\bibitem{tgi} Hoyland, R., Aguiar-Gonz{\'a}lez, M., G{\'e}nova-Santos, R., et al.\ 2014, \procspie, 9153, 915332 
\bibitem{kamionkowsky97} Kamionkowski, M.,
Kosowsky, A., \& Stebbins, A.\ 1997, \prd, 55, 7368
\bibitem{quijote14} L{\'o}pez-Caniego, M., Rebolo, R., Aguiar, M., et al.\ 2014, arXiv:1401.4690 
\bibitem{lopez11} L{\'o}pez-Caraballo, C.~H., Rubi{\~n}o-Mart{\'{\i}}n, J.~A., Rebolo, R., \& G{\'e}nova-Santos, R.\ 2011, \apj, 729, 25 
\bibitem{polarbear14} The Polarbear Collaboration: P.~A.~R.~Ade, 
Akiba, Y., Anthony, A.~E., et al.\ 2014, \apj, 794, 171
\bibitem{cpp1} Planck Collaboration.
Planck 2013 Resuls I, 2014a, \aap, 571, AA1
\bibitem{pip30} Planck Collaboration.
Planck Intermediate Results XXX, 2014b, arXiv:1409.5738
\bibitem{quiet12} QUIET
Collaboration, Araujo, D., Bischoff, C., et al.\ 2012, \apj, 760, 145
\bibitem{quijote12} Rubi{\~n}o-Mart{\'{\i}}n, J.~A., Rebolo, R., Aguiar, M., et al.\ 2012, 
\procspie, 8444, 84442Y 
\bibitem{zaldarriaga97} Zaldarriaga, M., \& Seljak, U.\ 1997, \prd, 55, 1830
\end{thebibliography}
\end{document}